\documentclass{emulateapj}
\usepackage{epsfig}

\slugcomment{Accepted for publication by the Astrophysical Journal Letters}

\shorttitle{UY Vol in Quiescence}
\shortauthors{Hynes \& Jones}

\begin{document}

\title{The Quiescent Optical and Infrared Counterpart to EXO~0748--676 = UY~Vol}

\author{Robert I. Hynes\altaffilmark{1} and Erica D. Jones}
\affil{Louisiana State University, Department of Physics and Astronomy, 202 Nicholson Hall, Tower Drive, Baton Rouge, Louisiana 70803}

\altaffiltext{1}{rih@phys.lsu.edu}

\begin{abstract}
  We present optical and infrared photometry of the low-mass X-ray
  binary EXO~0748--676 in quiescence for the first time in 24 years
  since it became X-ray active in 1985.  We find the counterpart at
  average magnitudes of $R=22.4$ and $J=21.3$.  We monitored the
  source approximately nightly through 2008 November to 2009 January.
  During this time there was considerable night-to-night optical
  variability but no long term trends were apparent.  The
  night-to-night variability reveals a periodicity of
  $P=0.159331\pm0.000012$\,d, consistent with the X-ray orbital period
  to within 0.01\,\%.  This indicates that the quiescent optical
  modulation is indeed orbital in nature rather than a superhump.
  Interestingly, the modulation remains single-peaked with a deep
  minimum coincident with the times of X-ray eclipse, and there is no
  indication of a double-peaked ellipsoidal modulation.  This
  indicates that even in `quiescence' emission from the accretion disk
  and/or X-ray heated inner face of the companion star dominate the
  optical emission, and implies that obtaining an accurate dynamical
  mass estimate in quiescence will be challenging.
\end{abstract}

\keywords{stars: binaries: eclipsing ; stars: individual: UY~Vol ; X-rays: binaries}

\section{Introduction}

The low-mass X-ray binary (LMXB) EXO~0748--676 was discovered in 1985
as a transient X-ray source \citep{Parmar:1985a} and rapidly
associated with an optical counterpart, UY~Vol \citep{Wade:1985a}.
Unlike most X-ray transients, however, it did not decay back to a
quiescent state, but remained active and for the last two decades has
been considered part of the persistent LMXB population.  In 2008
August--September, \citet{Wolff:2008a} found EXO~0748--676 unusually
faint, at a factor of two below its typical {\it RXTE} brightness and
suggested that a transition to quiescence might finally be impending.
In late September {\it Swift} found the source at a low luminosity
consistent with quiescence and in early October {\it RXTE} failed to
detect the source at all, confirming that it had dropped to quiescence
\citep{Wolff:2008b}.  \citet{Degenaar:2009a} presented five {\it
  Swift} and two {\it Chandra} observations spanning late September to
early November, including the {\it Swift} data previously reported by
\citet{Wolff:2008b}.  They found the source declining during this
period from $16\times10^{33}$\,erg\,s$^{-1}$ to
$8.3\times10^{33}$\,erg\,s$^{-1}$ with the spectrum dominated by
thermal emission from the neutron star surface.

The optical counterpart, UY~Vol, was originally discovered around 17th magnitude
\citep{Wade:1985a}.  Subsequent studies found pronounced orbital
modulation arising from a combination of eclipses of the accretion disk
and X-ray heating of the companion star.  The brightness spanned $17.7
> B > 16.9$ \citep{Crampton:1986a,Schmidtke:1987a} with $\sim0.2$\,mag
variations between epochs, and $17.8 > V > 17.1$
\citep{vanParadijs:1988a}.  This brightness was typical of later
observations.  By 2008 October, the source had
substantially faded to $R\sim22$ \citep{Hynes:2008a,Torres:2008a}
reinforcing the conclusion that it was now quiescent.

EXO~0748--676 is an intriguing object for a number of reasons.  It is
unusual among neutron star LMXBs in being a quasi-persistent source
for which we have observed the entire period of activity and
furthermore have estimates of the pre-eruption luminosity
\citep{Garcia:1999a}.  This makes it a fascinating system in which to
study the post-eruption cooling curve and investigate the physics of
neutron star interiors \citep{Degenaar:2009a}.

It also has an optimal inclination angle yielding total X-ray eclipses
of the neutron star without being an accretion disk corona source,
shows periodic X-ray dips, and also exhibits type I X-ray bursts
\citep{Parmar:1986a,Gottwald:1986a}. The possible detection of
gravitationally red-shifted absorption lines during X-ray bursts
offered tantalizing prospects for constraining the neutron star
equation of state \citep{Cottam:2002a} although this detection could
not be reproduced in subsequent observations \citep{Cottam:2008a}.
Independently, \citet{Ozel:2006a} argued that soft equations of state
can be ruled out based on observations of other characteristics of
X-ray bursts in EXO~0748--676.

Since our optical observation at the end of 2008 October
\citep{Hynes:2008a}, we have been following the source nightly in the
optical.  We report here on the optical behavior in quiescence over
the following three months.

\section{Observations}

\begin{figure}
\begin{center}
\epsfig{width=2.4in,file=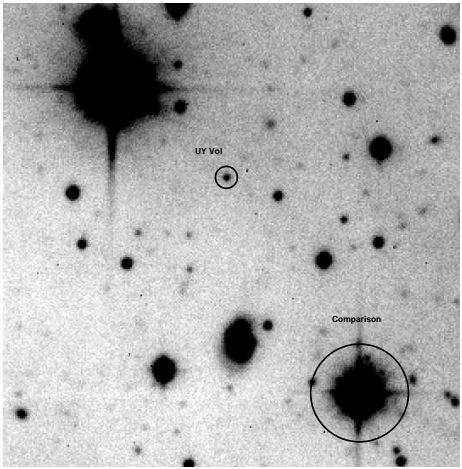}
\end{center}
\caption{Quiescent $R$ band finding chart of UY~Vol based on stacking
  all of our images.  The field of view is 100\arcsec\ square, with
  north at the top and east to the left.}
\label{ImageFig}
\end{figure}

Optical and infrared photometry of UY~Vol were obtained approximately
nightly from 2008 October 28 to 2009 February 5 using Andicam on the
SMARTS 1.3\,m telescope.  Each night, 3 450\,s $R$ band images were
taken together with 20 50\,s $J$ band images.  During this period data
were obtained on 71 out of 101 nights.  We show our combined optical
image in Fig.~\ref{ImageFig}.

Optical data were supplied with pipeline reductions applied and these
were satisfactory.  The three images from each night were aligned and
combined to produce a single average before performing aperture
photometry on UY~Vol and a comparison star (both shown in
Fig.~\ref{ImageFig}).  All photometry was done differentially relative
to the comparison star using standard IRAF\footnote{IRAF is
  distributed by the National Optical Astronomy Observatories, which
  are operated by the Association of Universities for Research in
  Astronomy, Inc., under cooperative agreement with the National
  Science Foundation.} techniques, and this star was then calibrated
separately relative to standard stars A, C, and D in the field of
T~Phe observed on multiple photometric nights \citep{Landolt:1992a}.
Our estimate of the calibrated magnitude of the comparison is
$R=14.66\pm0.03$.  The uncertainty quoted is the night-to-night
standard deviation.  Systematic errors may be larger as color
corrections were not possible since only $R$ band observations were
performed.

\begin{figure}
\epsfig{width=3.4in,file=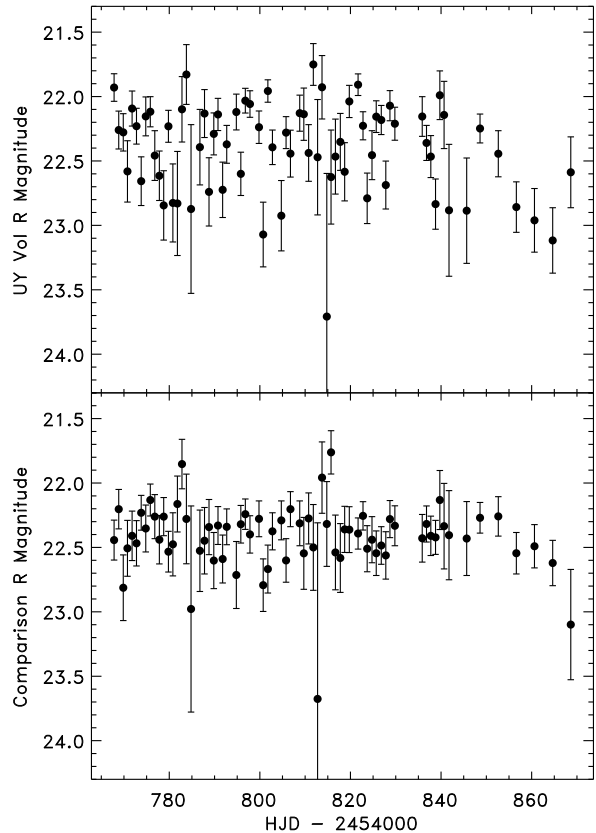}
\caption{Long-term SMARTS 1.3\,m $R$ band lightcurve.  The enhanced
  scatter around days 785, 815, and 845 correspond to full moon.  The lower
  panel shows a nearby comparison star of almost the same brightness;
  UY~Vol shows noticeably more variability.}
\label{LongLCFig}
\end{figure}

Our deduced average magnitude for UY~Vol is $R=22.39\pm0.04$.  This is
the formal error on the mean and systematic uncertainties in the
calibration may be larger.  We show the long-term lightcurve in
Fig.~\ref{LongLCFig}.  Considerable night-to-night variability is
present but there is no obvious long-term trend.  We will quantify
this statement after removing some of the intrinsic variability in
Section~\ref{LightcurveSection}.  To verify the significance of the
variability seen we also show a lightcurve for another nearby star at
$R=22.42\pm0.04$.  The standard deviation of the individual UY~Vol
data is 0.36\,mag, while that for the non-variable star is 0.26\,mag.

IR data were obtained in a 7-point dither pattern.  $3\times50$\,s images
were taken at 6 of the 7 positions, and 2 at the other.  For each
night, a sky image derived from the median of the dithered images was
subtracted, and flat-fields were applied.  We excluded images with the
highest sky values and those with a sky value significantly deviating
from the nightly mean to minimize residuals in the background
subtraction.  We then filtered the remaining images based on
visibility of the faintest stars in the field.  Our final combination
of these best images used 422 individual frames, all of which had been
individually checked.  The target is marginally detected in this
combined IR image at a position consistent with that measured from
optical images.  Photometry relative to several 2MASS stars in the
field yields $J=21.3\pm0.2$.  At this level, we caution that
systematic errors in background subtraction are likely to be larger
than the formal statistical error quoted.

\section{Period Search}

\begin{figure}
\epsfig{angle=90,width=3.4in,file=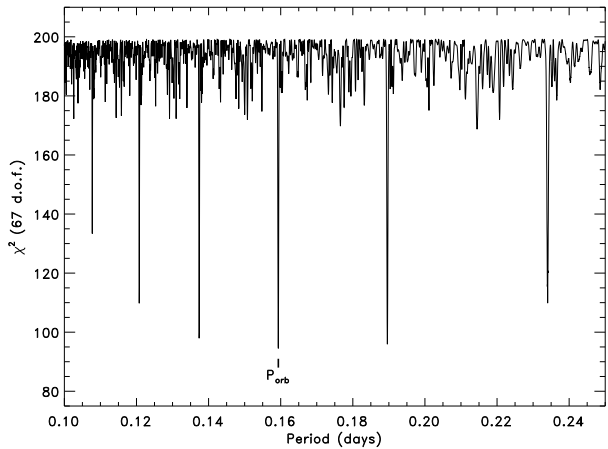}
\caption{$\chi^2$ as a function of trial periods from a sine-wave fit
  to the data.  The orbital period is marked.  The other prominent
  minima are consistent with one-day aliases of the orbital period.
  To avoid under-sampling issues, the fit was performed on a much finer
  period grid than shown, with each plotted point representing the
  minimum $\chi^2$ within its bin (rather than the average).}
\label{SineFitFig}
\end{figure}

It was immediately apparent that the data appeared consistent with
modulation on the published orbital period of 3.82\,hrs = 0.1593\,d
\citep{Wolff:2002a} with a single-humped modulation.  To verify this we
fitted the full dataset with a sinusoidal modulation of variable
period, allowing the phasing, amplitude and mean brightness to vary
freely.  We find several strong minima in the 0.10--0.25\,d period
range (Fig.~\ref{SineFitFig}).  One of these is consistent with the
orbital period, and the others are consistent with one-day aliases of
the orbital period, as expected given our once-per-day sampling.  No
significant minima are seen other than these aliases.

Choosing the alias corresponding to the X-ray period, we derive an
optical period of $P=0.159331\pm0.000012$\,d.  The $1-\sigma$
uncertainty quoted corresponds to the range of periods with which
$\chi^2 \leq \chi^2_{min}+1$. This period is consistent with
the secure X-ray orbital period of 0.159338\,d to within errors.  The
uncertainty is $\sim0.01$\,\%\ of the orbital period, so the optical
period is indeed orbital not significantly longer as would be expected
from a superhump modulation.

\section{Lightcurves}
\label{LightcurveSection}

\begin{figure}[t]
\epsfig{angle=90,width=3.4in,file=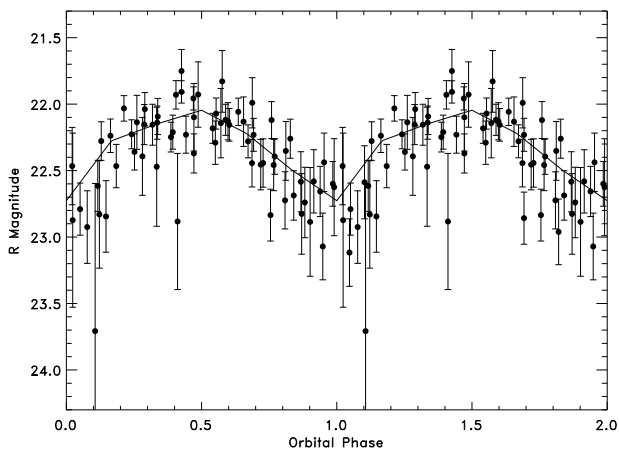}
\caption{Optical data folded on the X-ray orbital ephemeris.  The data
are plotted twice for continuity.  The solid line is a phase-binned version of the light curve to better show the average shape.}
\label{OrbitalLCFig}
\end{figure}

Since the dominant photometric variations are orbital, we fold the
data on the X-ray orbital ephemeris of \citet{Wolff:2002a} and show
the folded data in Fig.~\ref{OrbitalLCFig}.  The orbital modulation is
very apparent, with a minimum at phase 0.0 and a maximum at phase 0.5.
The lightcurve is rather similar to those seen in brighter states
\citep{Crampton:1986a,Schmidtke:1987a,vanParadijs:1988a} and
attributed there to a combination of eclipses of a hot accretion disk
and irradiation of the inner face of the companion star.  The
amplitude of the modulation is about the same as seen earlier.

We use the folded lightcurve to remove the orbital modulation and
better search for long-term trends.  In particular, we expect fading
if the optical light is due to heating by the cooling neutron star.
To do this we construct a coarsely phase-binned version of the
lightcurve, shown plotted over the data in
Fig.~\ref{OrbitalLCFig}. After removing the orbital modulation, no
long-term trend emerges and a formal fit yields a gradient of less
than 0.005\,mag per day, and consistent with zero.

\section{Discussion}

The lower range of the observed modulation provides an upper-limit
to the unheated brightness of the companion star.
\citet{MunozDarias:2009a} discuss the system parameters and nature of
the companion star based on outburst data.  They argue that the
companion mass is $0.16{\rm M}_{\odot}\leq M_2 \leq 0.42{\rm
  M}_{\odot}$, with the upper-limit corresponding to a main-sequence
mass donor and lower masses to somewhat evolved stars.  The
upper-limit suggests an M2V classification, for which we expect
$R=22.8$ at 7.7\,kpc \citep{Cox:2000a}.  The 7.7\,kpc distance was
estimated from a photospheric radius expansion burst by
\citet{Wolff:2005a} assuming a 1.4\,M$_{\odot}$ neutron star burning
helium rich material.  Assuming a heavier neutron star or hydrogen
rich material will both reduce the distance and hence increase the
expected brightness of the companion.  The lower range of our observed
modulation is approximately consistent with this brightness.  The
evolved case would be smaller, due to the smaller mass ratio and hence
smaller Roche lobe, and is likely to be cooler, so would be at
$R>22.8$. Given uncertainties in the distance, the main-sequence case
cannot be ruled out based on the current brightness of the source,
although could be if it fades below this level.

The modulation seen is clearly orbital in origin.  The period agrees
with the X-ray one to within 0.01\,\%, and the minimum coincides with
the minimum of the X-ray ephemeris.  This rules out superhumps as a
possible alternative modulation mechanism.  Superhumps may be common
in LMXBs \citep{Haswell:2001a} and have been seen in the black hole
system XTE~J1118+480 near quiescence \citep{Zurita:2002a}.  The
superhump period should exceed the orbital period by a few percent,
however, and this is clearly not the case.  

The modulation is single-peaked with a pronounced minimum near phase
zero.  It actually looks rather similar to those observed in bright
states, especially in the data of \citet{Schmidtke:1987a} where the
peak appears slightly skewed to phases earlier than 0.5.  Even the
amplitude is similar to the 0.5--0.6\,mag amplitude seen in outburst.
This suggests that the lightcurve is still dominated by X-ray heating
of the companion and/or eclipses of the accretion disk.  

The extent of X-ray heating of the companion star can readily be
estimated.  For a period of 3.82\,hrs and assuming a typical neutron
star mass, 1.4\,M$_{\odot}$, and a typical LMXB mass ratio of
one-third, we expect a binary separation $a\sim 10^{11}$\,cm.
Assuming an isotropic X-ray luminosity $L_{\rm
  X}=8.3\times10^{33}$\,erg\,s$^{-1}$ \citep{Degenaar:2009a}, we then
deduce an X-ray flux at the companion star of $f_{\rm
  X}=6\times10^{10}$\,erg\,cm$^{-2}$\,s$^{-1}$ and an irradiation
temperature of $T_{\rm irr} \simeq 6000$\,K for normal incidence.  Of
course, the true geometry does not provide normal incidence, and
thermal reprocessing of X-rays is likely to be less than 100\,\%\
efficient, but it would still be reasonable to expect an irradiation
temperature above 5000\,K over a significant area of the inner face of
the companion, well in excess of its likely photospheric temperature.
It is therefore plausible that X-ray heating (by the cooling neutron
star) could still dominate the lightcurve as seems to be observed,
though this is likely marginal at this point, and the unheated
portions of the photosphere may be contributing non-negligible flux.
We note that the deduced color, $R-J=1.1$ also independently suggests
an average effective temperature for the quiescent emission around
5000\,K, consistent with these calculations.

While the heating of the companion star can account for the
observations, some heating of the accretion disk may be occuring as
well.  This is harder to quantify as it depends sensitively on the
disk geometry.  In general, we would expect irradiation of the disk to
be less important in quiescence.  Cooler disks should be less
vertically extended so will intercept less X-ray flux.  This means
they will also shield the companion star less, so that the companion
should receive a larger fraction of the X-ray luminosity than in
outburst.  The similarity of the lightcurve shape and amplitude to
that in outburst certainly is suggestive that the disk may still be
contributing, but heating of the companion alone is sufficient to
explain the observations, so we cannot make a firmer statement on the
extent of the disk contribution.

A single-humped irradiation dominated lightcurve is not unprecedented
in quiescent objects.  This is what is seen in the optically bright
accreting milli-second pulsars in quiescence: SAX~J1808.4--3658
\citep{Burderi:2003a,Campana:2004a} and IGR~J00291+5934
\citep{DAvanzo:2007a,Jonker:2008a}.  In those systems, however, the
observed X-ray luminosity was two orders of magnitude too low to
explain the observed heating, and instead a turned-on pulsar wind was
invoked.  In EXO~0748--676 thermal emission from the cooling neutron
star is sufficient to explain the modest level of heating of the
companion that is needed.

While the neutron star appears to still be cooling
\citep{Degenaar:2009a}, pre-eruption X-ray observations suggest that
it may already be close to the true-quiescent X-ray level,
corresponding to the crust quickly returning to thermal equilibrium with an
unusually hot core \citep{Garcia:1999a,Degenaar:2009a}.  In this case,
we would not expect the optical counterpart to dim significantly more,
and that the X-ray heating may remain a persistent feature in the
quiescent lightcurve.  This is supported by our lack of apparent
long-term decay. 

This means there may be limited prospect for measuring ellipsoidal
variations.  Fortunately, since UY~Vol is an eclipsing system we
already have good constraints on the binary inclination without
ellipsoidal variations.  Obtaining a radial velocity curve is in
principle still possible, as the X-ray heating is at a low level and
is likely not completely overwhelming photospheric emission.  Other
X-ray active systems have yielded photospheric radial velocity curves,
e.g.\ V395~Car \citep{Shahbaz:1999a}, although concerns are then
raised about whether the photospheric absorption lines are present
across the whole companion star surface and hence whether the radial
velocity curve determined truly traces the motion of the companion star
center of mass.  In any case, with a brightness of $22 < R < 23$, UY~Vol
will be an extremely challenging target for radial velocity studies.

\section{Conclusions}

We have obtained optical photometry of EXO~0748--676 = UY~Vol since
its X-ray decline to quiescence in late 2008.  We find an average
brightness of $R=22.4$, with substantial intrinsic variability.  The
variations are orbital in origin and define a single-humped lightcurve
similar to that seen in outburst.  This suggests that X-ray heating by
the cooling neutron star is still dominating the optical emission and
that obtaining a reliable dynamical mass estimate may be challenging
for this object.

\acknowledgments

We are grateful to the staff of the SMARTS Consortium for performing
these observations superbly.  This work was supported by
NASA/Louisiana Board of Regents grant NNX07AT62A/LEQSF(2007-10)
Phase3-02.  This work has made use of the NASA Astrophysics Data
System Abstract Service.

\clearpage

\end{document}